\documentclass[conference]{IEEEtran}

\usepackage{url}
\usepackage{cite}
\usepackage{amsmath,amssymb,amsfonts}
\usepackage{algorithmic}
\usepackage{graphicx}
\usepackage{textcomp}
\usepackage{xcolor}
\usepackage{hyperref}
\usepackage{enumitem}
\usepackage{array}
\usepackage{multirow}
\usepackage{makecell}
\usepackage{adjustbox}
\usepackage{listings}
\usepackage{tcolorbox}
\tcbuselibrary{listingsutf8}
\usepackage{stfloats}

\begin{document}

\title{Secure Development of a Hooking-Based Deception Framework Against Keylogging Techniques}

\author{
\IEEEauthorblockN{Md Sajidul Islam Sajid}
\IEEEauthorblockA{\textit{Computer and Information Sciences} \\
\textit{Towson University} \\
Towson, USA \\
msajid@towson.edu}
\and
\IEEEauthorblockN{Shihab Ahmed}
\IEEEauthorblockA{\textit{Computer and Information Sciences} \\
\textit{Towson University} \\
Towson, USA \\
sahmed23@students.towson.edu}
\and
\IEEEauthorblockN{Ryan Sosnoski}
\IEEEauthorblockA{\textit{Computer and Information Sciences} \\
\textit{Towson University} \\
Towson, USA \\
rsosno1@students.towson.edu}
}

\maketitle
\IEEEpeerreviewmaketitle

\begin{abstract}
Keyloggers remain a serious threat in modern cybersecurity, silently capturing user keystrokes to steal credentials and sensitive information. Traditional defenses focus mainly on detection and removal, which can halt malicious activity but do little to engage or mislead adversaries. In this paper, we present a deception framework that leverages API hooking to intercept input-related API calls invoked by keyloggers at runtime and inject realistic decoy keystrokes. A core challenge, however, lies in the increasing adoption of anti-hooking techniques by advanced keyloggers. Anti-hooking strategies allow malware to bypass or detect instrumentation. To counter this, we introduce a hardened hooking layer that detects tampering and rapidly reinstates disrupted hooks, ensuring continuity of deception. We evaluate our framework against a custom-built ``super keylogger” incorporating multiple evasion strategies, as well as 50 real-world malware samples spanning ten prominent keylogger families. Experimental results demonstrate that our system successfully resists sophisticated bypass attempts, maintains operational stealth, and reliably deceives attackers by feeding them decoys. The system operates with negligible performance overhead and no observable impact on user experience. Our findings show that resilient, runtime deception can play a practical and robust role in confronting advanced threats.    

\end{abstract}

\section{Introduction}\label{intro}


Keyloggers remain a persistent threat, operating silently in the background to capture user keystrokes and extract sensitive or financial information, often before the data is exfiltrated to the attacker. The presence of keyloggers allow the attacker to access information early, at the interface between human input and system processing. Most defensive strategies focus on identifying and terminating keyloggers through static signatures, behavioral heuristics, or anomaly detection. While these methods help contain potential harm, they also eliminate the opportunity to gather intelligence on the adversary or modify their behaviors through deception.


Cyber deception provides a powerful complement to traditional detection-based defenses. Rather than simply identifying threats, deception seeks to delay, confuse, and mislead adversaries by inserting false or misleading information into their operational environment. This approach disrupts the attacker's decision-making process, contaminates any exfiltrated data, and undermines their ability to act effectively during exploitation. In the context of keylogging, deception enables defenders to corrupt captured keystrokes with disinformation, turning stolen input into misleading artifacts that can support attacker attribution or strengthen existing detection mechanisms. To implement deception, we utilize API hooking, a dynamic instrumentation technique that intercepts and modifies function calls during execution. In contrast to heavier approaches such as virtualization or full-system sandboxing, API hooking offers fine-grained control with minimal performance overhead. It also enables lightweight deployment and provides real-time visibility into the use of critical API functions. In our framework, we target input-related APIs that are exploited by keyloggers in order to inject decoy data, simulate keystroke activity, and monitor attacker behavior. However, modern adversaries are increasingly equipped with advanced techniques to detect and evade instrumentation. These evasion strategies are specifically designed to neutralize defenses based on API hooking and can render a naive deception approach ineffective. A deception system that does not account for such countermeasures may fail to operate reliably in adversarial environments. Our work addresses this challenge by securing both the deceptive outputs and the resilience of the underlying hooking infrastructure. The framework is designed to detect tampering, recover from manipulation, and remain operational even when subjected to active evasion attempts. This combined focus on deception and robustness distinguishes our approach from prior efforts.

Several prior works have explored keylogger detection, API monitoring, and honeypot-style deception. For example, Nasaka et al.\cite{nasaka2011keystroke} monitored keyboard APIs to detect keystroke loggers; Al-Husainy\cite{al2008detecting} intercepted API calls for bot detection; and HookTracer~\cite{case2020hooktracer} used memory forensics to identify hooks. Other studies, such as~\cite{muthumanickam2015effective}, focused on detecting hook tampering through Import Address Table (IAT) and inline hook monitoring. However, these efforts primarily focus on detection rather than deception. While works such as \cite{sajid2021soda, sajid2023symbsoda, islam2021chimera, ahmed2025spade, islam2025randecepter} demonstrated the use of API hooking to achieve deception, they do not address the resilience of the hooking layer against modern anti-hooking techniques. Sophisticated adversaries can bypass such defenses using methods like prologue restoration, dynamic API resolution, encrypted input channels, or shadow DLL loading.
However, these works generally focus on either detection or static deception and are not designed to persist under active tampering.

In contrast, our work presents a deception framework that not only injects decoy inputs but is also hardened against runtime evasion. It actively detects and responds to tampering attempts, preserving both the continuity of deception and the stealth of its instrumentation. To the best of our knowledge, this is the first deception framework based on API hooking that includes runtime resilience against advanced keyloggers equipped with anti-hooking capabilities. Overall, this paper makes the following contributions:

\begin{itemize}
  \setlength\itemsep{0.3em}
  \item We conduct a systematic analysis of userland keylogging techniques and identify the critical input-related APIs most frequently exploited for keystroke capture.
  
  \item We examine four representative anti-hooking strategies commonly used in malware, academic proposals, and public tooling, and evaluate their impact on user-mode API hooking.

  \item We design and implement a secure deception framework built on EasyHook and Microsoft Detours, enabling realistic decoy injection while incorporating layered defenses to maintain hook integrity against tampering.

  \item We evaluate the framework against both a custom multi-vector keylogger with advanced anti-hooking and real-world malware samples. The results demonstrate that our system detects and mitigates evasion attempts in real time, reliably delivers decoy inputs with a minimal performance overhead.
\end{itemize}


The remainder of this paper is organized as follows. Section~\ref{background} provides background on keylogging and anti-hooking techniques, research questions, and the threat model. Section~\ref{Framework} details the design of our deception framework, including architecture and tamper resilience. Section~\ref{eval} presents implementation and evaluations against real-world and simulated keyloggers. Section~\ref{Related} reviews related work. Finally, Section~\ref{Conclusion} summarizes our findings, discusses limitations, and concludes with directions for future research.

\section{Background, Research Questions and Threat Model}\label{background}

\subsection{Keyloggers}


Keyloggers are surveillance malware that capture user keystrokes to extract sensitive input such as credentials, financial information, or private messages. The stolen data is exfiltrated to attacker-controlled servers for identity theft, fraud, or unauthorized access. While hardware-based variants exist, this work focuses on user-space software keyloggers. These typically exploit OS APIs by polling keystates, registering hooks, monitoring message queues, or intercepting clipboard and form data. Table~\ref{tab:keyloggers} summarizes common techniques and APIs identified through our literature review. A complete mapping of surveyed methods and API usage is available in the anonymized supplementary repository~\cite{keylogger_techniques}.

\begin{table}[h]
\vspace{-2mm}
\centering
\caption{Userland Keylogger Techniques and Associated APIs}
\label{tab:keyloggers}
\scriptsize
\begin{tabular}{|p{2.2cm}|p{5.5cm}|}
\hline
\textbf{Technique} & \textbf{Common Userland APIs} \\
\hline
Keyboard Hooking & \texttt{SetWindowsHookEx (WH\_KEYBOARD\_LL)}, \texttt{UnhookWindowsHookEx} \\
\hline
Polling Keystates & \texttt{GetAsyncKeyState}, \texttt{GetKeyState} \\
\hline
Message Queue Interception & \texttt{PeekMessage}, \texttt{GetMessage} \\
\hline
Clipboard Monitoring & \texttt{OpenClipboard}, \texttt{GetClipboardData} \\
\hline
Form Grabbing & \texttt{HttpSendRequest}, \texttt{InternetWriteFile}, \texttt{WSASend} \\
\hline
Screen Logging & \texttt{BitBlt}, \texttt{PrintWindow}, \texttt{GetDC} \\
\hline
Mouse Tracking & \texttt{SetWindowsHookEx (WH\_MOUSE\_LL)}, \texttt{GetCursorPos} \\
\hline
Browser Script Injection & \texttt{onKeyDown}, \texttt{onKeyPress}, \texttt{onKeyUp} (JavaScript DOM APIs) \\
\hline
Raw Input Capture & \texttt{RegisterRawInputDevices}, \texttt{GetRawInputData} \\
\hline
Process Injection & \texttt{WriteProcessMemory}, \texttt{CreateRemoteThread} \\
\hline
\end{tabular}
\vspace{-15pt}
\end{table}


\subsection{Deception Systems}


Cyber deception misleads attackers by injecting false information or deploying decoys to expose malicious behavior and increase uncertainty. While commonly used in network security and access control, it is also effective against keyloggers. For keyloggers, deception involves simulating user activity and injecting fake keystrokes to corrupt captured data. For example, \cite{simms2017keylogger} used a decoy keyboard to confuse keyloggers, and \cite{wazid2013framework} proposed a honeypot that simulates interaction and detects unauthorized access. Unlike detection methods that terminate threats, deception enables controlled engagement to gather intelligence, trigger behavior signatures, and prolong adversary exposure.

\subsection{API Hooking}
API hooking is a widely used technique in system security and malware analysis for monitoring and modifying application behavior at runtime. It intercepts system or library calls to provide visibility and control, enabling sandboxes, antivirus tools, and debuggers to detect suspicious activity and enforce policies. Its dynamic, non-invasive nature supports deployment on live systems without kernel access. In cyber deception, API hooking enables real-time intervention as malware attempts to collect sensitive data. It is particularly effective against userland keyloggers that rely on APIs like \texttt{GetAsyncKeyState}, \texttt{SetWindowsHookEx}, and \texttt{ReadConsoleInput}. By intercepting these calls, defenders can inject false inputs, corrupt logs, or trigger early exfiltration. Unlike static deception, API-level deception enables dynamic, policy-driven interaction. However, advanced malware may use anti-hooking tactics to detect or disable hooks, requiring defensive deployments to emphasize stealth, integrity, and tamper resilience.

\subsection{Motivation, Anti-Hooking, and Research Questions}\label{Motives}

A core challenge for hooking-based deception is resilience against anti-hooking techniques employed by advanced keyloggers. Modern malware can detect, bypass, or disable userland API hooks, threatening defensive frameworks built using prominent hooking libraries such as \textbf{EasyHook}\cite{easyhook} and \textbf{MS Detours}\cite{microsoft_detours}. To ensure robust deception, our framework explicitly counters four widely observed anti-hooking strategies~\cite{smokeloader2,0x00sec-defeating-hooks, cano_reloadlibrary, apostolopoulos2021resurrecting}: (i) .text section restoration to remove inline hooks, (ii) IAT rebinding to clean DLL copies, (iii) alternate DLL loading to evade instrumentation, and (iv) memory scanning to identify hook signatures. In these techniques, adversaries overwrite trampoline-modified code regions with clean instruction bytes, rebind imported function pointers in the IAT, load alternate DLLs (e.g., ntdll.dll), or scan memory for hook patterns to avoid instrumented functions. Such actions undermine any defense that assumes persistent control at the userland API boundary. Without effective tamper detection and mitigation, hooking-based deception may easily fail. Therefore, addressing these evasion strategies is essential for building an effective and resilient deception system. To guide the design and evaluation of our framework, we pose the following research questions:

\begin{itemize}
  \setlength\itemsep{0.3em}
  \item \textbf{RQ1:} Which API surfaces do userland keyloggers most frequently use, and what opportunities do they offer for deploying deception?
  \item \textbf{RQ2:} What anti-hooking techniques are employed by keyloggers to bypass userland monitoring, and how can a deception framework be designed to preserve its operational integrity in adversarial environments?
\end{itemize}

\subsection{Threat Model}


We consider user-mode keyloggers that operate with standard user privileges on a compromised system to capture sensitive inputs such as passwords or form data. These keyloggers may rely on common techniques including keyboard polling, input hooks, GUI message monitoring, clipboard access, or form grabbing. The attacker is assumed to be sophisticated, using anti-hooking tactics such as unhooking prologues, DLL shadow loading, dynamic API resolution, and memory scanning. Kernel-level and hardware keyloggers are out of scope. Our framework is not intended for malware detection. Instead, it becomes active after an external detection mechanism identifies a process as malicious. Such mechanisms may include an Endpoint Detection and Response (EDR) system such as~\cite{arfeen2021endpoint}, intervention by a human analyst or prior detection systems~\cite{ortolani2010bait, ahmed2019key, pillai2019modified}. Once flagged, the deception module is injected into the process and operates at the same privilege level. The defender hooks input-related APIs, assuming OS and library integrity (e.g., EasyHook, Detours), and no interference from other endpoint security tools. The framework aims to \textbf{(1) prevent data leakage} by feeding keyloggers false inputs, \textbf{(2) preserve stealth} so attackers remain unaware of deception, and \textbf{(3) resist evasion and tampering} by detecting and recovering from common anti-hooking techniques to maintain the integrity of its instrumentation.

\section{Framework Design}\label{Framework}

\begin{figure}[!ht]
    \centering
    \includegraphics[width=0.49\textwidth] {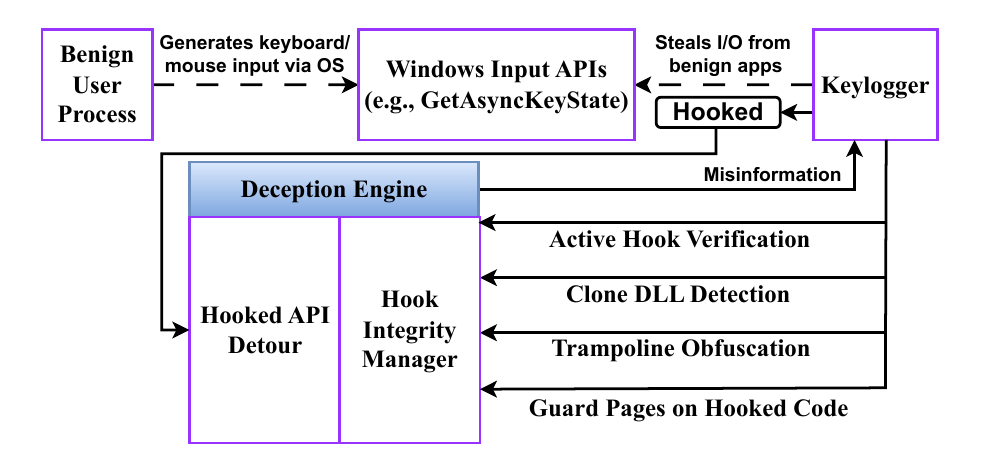}
    \vspace{-0.75cm}
    \caption{System Architecture of the Deception Framework.}
    \label{fig:system_architecture}
\vspace{-2pt}
\end{figure}

Our deception framework intercepts keylogger activity in real time, injects realistic but misleading inputs, and maintains its own integrity under attacker tampering. We build on EasyHook and Microsoft Detours—two widely used API hooking libraries—to achieve lightweight, runtime interception of relevant APIs. The architecture consists of three core components: a \textbf{Hooking Layer} that implants inline hooks at chosen input APIs, a \textbf{Deception Engine} that decides how to modify or fabricate data returned to the keylogger, and a \textbf{Hook Integrity Manager} that monitors and preserves the hooks against removal attempts. In essence, the Hooking Layer provides interception, the Deception Engine injects decoys, and the Integrity Manager ensures continued operation under adversarial conditions. Figure \ref{fig:system_architecture} provides the system architecture of the deception framework.

\subsection{Target API Coverage}\label{api_coverage}
We focused on the Windows platform, as more than ninety percent of new malware samples target Windows systems~\cite{malwarestatistics}. Its mature and well-documented input APIs make it especially attractive to attackers. Our implementation uses both EasyHook and Microsoft Detours. While Microsoft Detours is tailored specifically for Windows, EasyHook is platform-agnostic. Our system can be adapted to other platforms with equivalent interception capabilities, once the relevant input APIs are identified for those environments.

A critical design consideration in building our deception framework was identifying which API calls to intercept to effectively undermine keyloggers. To identify these crucial APIs, we conducted a threat-driven study of fifty (50) keylogger samples drawn from ten widely observed keylogger families: Agent Tesla, AppleSeed, RokRAT, gh0st RAT, WarzoneRAT, Astaroth, Micropsia, NightClub, Azorult, and Lokibot.  The samples were collected from publicly available malware repositories, including AnyRun and VirusTotal. For each family, we selected five representative binaries based on observed behavioral diversity and prevalence in recent campaigns. Each sample was dynamically analyzed using Cuckoo Sandbox, Rohitab API Monitor, and OllyDBG. We focused our efforts on the keylogging collection phase, as this stage reveals attacker intent most directly and offers a more tractable point for defensive intervention. Later stages such as file manipulation or network exfiltration often involve encryption or protocol obfuscation that makes deception less practical.

To identify the most relevant API calls used for gathering user input, we manually examined execution traces. Rather than relying on raw API call frequencies, we applied expert judgment to reconstruct the control flow and analyze the logic used by each malware sample to capture input. This allowed us to filter noise and highlight only those APIs that were functionally essential for keylogging. These categories, along with representative examples and associated keylogging strategies, are summarized in Table~\ref{tab:keyloggers}.  In particular, we concentrated on APIs that (i) directly expose user input, (ii) allow indirect inference of keystrokes through GUI behavior, or (iii) facilitate passive capture of input data before exfiltration. We identified four key categories of target APIs:


\textbf{Direct Input Acquisition APIs:} Functions that immediately return keyboard state, which are popular for straightforward keylogging. For example, many malware use polling calls like \texttt{GetAsyncKeyState} or \texttt{GetKeyState} within loops to detect keystrokes with minimal latency. Our Hooking Layer intercepts these calls and returns manipulated key states or injected noise. For instance, when a keylogger polls these APIs, our system may respond with a stream of plausible but fake keystrokes, effectively corrupting the logged data at its source and rendering the captured information unreliable..


\textbf{Window Message Inspection:} APIs such as \texttt{GetMessage} and \texttt{PeekMessage} enable malware to monitor the GUI message queue and infer user keystrokes based on intercepted window messages. Our analysis shows that keyloggers targeting graphical applications, particularly browsers and email clients, frequently uses this technique to extract typed input. To counter this behavior, our framework intercepts message-level activity and introduces deception. When keylogging is detected within sensitive applications, our hooks can selectively suppress or manipulate \texttt{WM\_KEYDOWN} and \texttt{WM\_CHAR} messages. For example, the framework may inject fake key events or alter character messages to produce misleading input traces in the keylogger’s log file.


\textbf{Hook-Based Capture APIs:} Some keyloggers rely on legitimate Windows APIs such as \texttt{SetWindowsHookEx} with the \texttt{WH\_KEYBOARD\_LL} flag to register global keyboard hooks and intercept keystrokes system-wide. Our framework counters this technique through two mechanisms. First, it can block untrusted processes from registering keyboard hooks by failing the hook installation or actively removing it. Second, if a hook is successfully registered, our system intercepts and overrides its callback logic. This allows us to prevent the keylogger from accessing real keystroke data and instead deliver fabricated input. These defenses are critical for mitigating threats that abuse legitimate OS-provided hooking capabilities.


\textbf{Auxiliary Data Collection:} In addition to capturing keystrokes directly, keyloggers often exploit supporting APIs to collect sensitive data or exfiltrate user input. Common examples include clipboard access via \texttt{OpenClipboard} and \texttt{GetClipboardData}, or intercepting form submissions using network functions like \texttt{HttpSendRequest} and \texttt{WSASend}. Our Hooking Layer monitors these APIs as well, allowing the Deception Engine to inject plausible decoy content. For instance, fake clipboard data can be returned to mislead malware accessing \texttt{GetClipboardData}, and fabricated credentials can be inserted into outbound transmissions. Since these APIs are typically more context-sensitive, hooks are selectively applied to preserve application stability while disrupting form-grabbing behavior. This categorization guided our API selection and prioritization: lower-level functions such as polling or clipboard access allow simple, low-overhead interception; GUI and hook-based methods provide richer context but require careful handling; and network-layer APIs, while demanding deeper inspection, offer strategic points to corrupt attacker objectives. Our framework balances these trade-offs, achieving broad coverage and effective deception with minimal performance overhead, as demonstrated in our evaluations.


\vspace{-1pt}

\subsection{Input Deception Policies}
Intercepting an API call is the initial step. The core functionality of the Deception Engine lies in determining how to respond to those intercepted calls. Since our framework is deployed only after a process has been externally flagged as suspicious, such as by an analyst, Endpoint Detection and Response (EDR) tool, or anti-malware system, it operates in a strictly \textbf{active mode}, applying deception policies immediately and consistently. 
This design avoids the need for runtime decision-making, behavioral triggers, or passive logging within the framework itself.

\textbf{Decoy Injection:} The hook returns synthetic data in place of the real input. This straightforward strategy directly feeds false information to the keylogger. For instance, imagine malware polling GetAsyncKeyState to capture a user’s password as they type. Under a Decoy Injection policy, when the user presses ``P", the keylogger might instead receive a code corresponding to, say, ``M", followed by a sequence of characters spelling out a fake password. In our prototype, we scripted decoy payloads such as ``Password123!" or realistic keyboard patterns to simulate user typing. If the attacker is exfiltrating keystrokes in real time or operating in an automated fashion, this tactic poisons their data entirely. We found Decoy Injection particularly effective against rapid key thieves, as it fills their logs with plausible-looking but incorrect credentials.

\textbf{Input Perturbation:} Rather than full fabrication, the engine makes slight, often random modifications to the legitimate input data. The goal is to degrade the fidelity of stolen data while remaining subtle. 
Our framework can, for example, flip the case of every third character the user types, introduce small delays in reporting keystrokes, or inject an occasional extra keystroke that would be benign (like an innocuous typo) from the user’s perspective. This way, the attacker’s logged keystroke sequences will not exactly match what the user actually entered. In one test scenario, we enabled a perturbation policy that would randomly swap two characters in a password being captured. The genuine login still succeeded (because the real keystrokes went through to the application unaltered), but the keylogger’s record contained the wrong characters, thwarting any later misuse of that password.

Input modification is configurable at deployment via a masking ratio (e.g., 30\% or 100\%), enabling partial perturbation of the keystroke buffer through randomized index selection. This reduces overhead while maintaining plausible deception, especially valuable under high-frequency polling, where full modification may introduce latency. To ensure efficiency, the system adopts a context-agnostic strategy that avoids semantic input analysis. While context-aware deception (e.g., leveraging NLP or LLMs) is feasible, it introduces additional complexity and cost, hence is left as future work.

\subsection{Hook Integrity and Tamper Resilience}\label{resistance}


While the Hooking Layer and Deception Engine mislead attackers, an advanced adversary may attempt to disable the framework by tampering the hooks. Hence, we prioritized hook integrity from the start, guided by real attacker behaviors. The Hook Integrity Manager implements layered defenses to detect and counter such interference. Hook Integrity Manager uses memory protection attributes such as \texttt{PAGE\_NOACCESS} or \texttt{PAGE\_GUARD} to safeguard its internal data and control logic. It executes as a privileged watchdog thread outside common scanning regions, reducing its exposure to tampering. These protections help isolate the Hook Integrity Manager from untrusted threads while maintaining operational transparency.


\textbf{Active Hook Verification:} EasyHook does not provide built-in support for verifying the integrity of installed hooks. To address this limitation, our framework implements custom logic to detect and recover from hook tampering in real time. After installing a hook using \texttt{LhInstallHook}, we capture a lightweight hash or byte-level snapshot of the target function’s prologue. A dedicated watchdog thread periodically compares the current memory content of each hook target with the expected detour layout. If an attacker attempts to unhook (e.g., restoring the original function bytes via \texttt{memcpy} or \texttt{WriteProcessMemory}), the tampering is detected and the Hook Integrity Manager immediately restores the hook. This self-healing process preserves the continuity and stealth of the deception engine. In our evaluation, tampering was typically corrected within milliseconds and did not introduce measurable performance degradation.


\textbf{Clone DLL Detection:} Some advanced malware attempt to evade userland hooks by loading fresh copies of common libraries such as \texttt{user32.dll} or \texttt{ntdll.dll} under alternate names or paths. By calling functions from these clean instances, the malware can bypass previously installed hooks. While effective against static hooking, our framework extends EasyHook with custom logic to detect and counter such evasion. Since EasyHook does not monitor module load events by default, we hook low-level loader functions, such as \texttt{LdrLoadDll}, within the target process. When a new module is loaded, our hook inspects its export table and code sections to identify clones of known system libraries. If matched, we dynamically apply the same set of hooks to the new module using \texttt{LhInstallHook} API. This clone-aware mechanism ensures continued interception even when malware attempts to establish an unmonitored API path through DLL shadowing. In our evaluation, this method successfully detected and re-hooked a dynamically loaded clean copy of \texttt{user32.dll}, neutralizing the bypass attempt in real time.


\textbf{Trampoline Obfuscation:} By default, EasyHook installs inline detours using common jump instructions such as \texttt{JMP rel32} (\texttt{E9}) or indirect jumps via pointers (e.g., \texttt{FF25}). These opcode patterns are widely recognized and easily identified through memory scans that target the prologues of commonly hooked functions. To enhance stealth, we implemented a custom trampoline generator that avoids these recognizable signatures. Our obfuscation logic constructs semantically equivalent but structurally different sequences, including \texttt{PUSH address; RET} and \texttt{MOV reg, address; JMP reg}. We further randomize the detour memory layout and insert benign padding instructions to disrupt signature-based scanning. These transformations are applied dynamically during hook installation, yielding a distinct hook fingerprint for each deployment. In evaluation, our obfuscation consistently evaded detection by simulated attacker tools that relied on signature matching. While this functionality extends beyond EasyHook’s native features, it significantly increases resistance to static memory analysis and reverse engineering.

\begin{table*}[h]
\centering
\footnotesize
\caption{Defensive Measures Implemented by the Hook Integrity Manager}
\label{tab:HIM}
\begin{tabular}{|p{2.0cm}|p{7.0cm}|p{3.5cm}|p{3.5cm}|}
\hline
\textbf{Defense Mechanism } & \textbf{Description} & \textbf{Tamper Detection Method} & \textbf{Reaction} \\
\hline
Active Hook Verification & Detects tampering by periodically comparing current memory at each hook site with the expected detour layout. Reinstalls hooks if tampering is detected. & Memory comparison with expected detour layout. & Reapply detour patch immediately. \\
\hline
Clone DLL Detection & Hooks module loader to detect duplicate DLLs (e.g., user32.dll). Applies the same set of hooks to newly loaded modules if a match is found. & Analysis of module characteristics (e.g., export table signatures or code sections) on load. & Apply hooks on duplicate modules \\
\hline
Trampoline Obfuscation & Replaces standard JMP instructions with alternate sequences to avoid signature recognition via memory scanning. & Avoidance of common jump signatures. & Avoid detection from memory scanners. \\
\hline
Guard Pages on Hooked Code & Modifies memory protection of target function’s page using PAGE\_GUARD or PAGE\_NOACCESS to catch tampering via access violations. & Access violation triggers via structured exception handler (SEH). & Intercept and restore hook; Optionally terminate thread. \\
\hline
\end{tabular}
\vspace{-15pt}
\end{table*}

\textbf{Guard Pages on Hooked Code:} As an additional layer of tamper detection, we implemented page-level memory protections to secure the code regions of hooked functions. After installing a hook, our framework modifies the memory protection of the corresponding page using \texttt{PAGE\_GUARD} or \texttt{PAGE\_NOACCESS}, depending on operating system support. Any unauthorized read or write access to this memory, such as an attempt to inspect or modify the function prologue, will trigger an access violation. The framework registers a structured exception handler (SEH) that intercepts these violations, verifies the integrity of the hook, and immediately restores it if tampering is detected. In stricter configurations, the attacking thread or process is terminated to prevent further manipulation attempts. This method effectively turns the hooked memory region into a tripwire, enabling real-time detection and response. In our controlled evaluation, this method successfully raised an \texttt{access violation} when a simulated keylogger attempted to overwrite the prologue of \texttt{GetAsyncKeyState}. The triggered exception allowed our framework to detect and restore the hook before any further bypass activity could occur. We note that this guard-based approach is not universally applicable, as some system DLL or shared memory regions may not reliably support guarded permissions due to OS level constraints. Nonetheless, in applicable scenarios, it provided a strong layer of hook resilience with minimal overhead. Table~\ref{tab:HIM} summarizes the defensive measures implemented by the Hook Integrity Manager. By integrating these defenses, the framework maintains resilience against tampering. The Hook Integrity Manager converts anti-hooking behaviors into observable events, allowing timely recovery with negligible performance overhead.

\section{Implementation and Evaluations}\label{eval}

\subsection{Prototype Implementation}
We tested our deception framework on virtualized Windows 10 and Windows 11 environments. Our deception framework was manually injected into target processes using EasyHook’s remote injection API. We tested it against both open-source and real-world malware samples. Evaluation metrics included CPU overhead, success of decoy injection, and resilience to tampering. Hook integrity was verified using memory hash comparisons and structured exception handler (SEH)-based validation. To validate the design of our deception framework, we developed two functionally equivalent prototypes using EasyHook and Microsoft Detours. Our goal was to determine whether the core ideas: the hook-based interception, decoy injection, and tamper resilience could be implemented effectively across different hooking libraries, each with their own engineering trade-offs.

The EasyHook-based prototype provided a high-level API, support for both 32-bit and 64-bit processes, and native .NET integration. These features enabled rapid development and iterative testing with minimal low-level engineering. In contrast, the Detours-based implementation offered fine-grained control over instruction patching. We extended it with custom logic for randomized trampoline generation and runtime hook validation to enhance stealth and robustness. While this version required more manual engineering effort, particularly for handling edge cases in 64-bit systems, it validated the generality and portability of our framework design. For clarity and consistency, our evaluation focuses on the EasyHook-based prototype, which served as the basis for case studies and performance testing. The complete source code for both variants is publicly available at~\cite{keylogger_techniques}.

\subsection{Case Study: Deceiving WarzoneRAT, a Real-World Polling-Based Keylogger}\label{evalB}
We deployed our framework against WarzoneRAT (a.k.a. Ave Maria RAT) \cite{warzonerat}, a widely documented remote access trojan with a built-in keylogger. We selected WarzoneRAT as a case study because it represents a common but effective class of keyloggers. Its polling-based approach, though simple, generates high API call volume, making it ideal for testing the stability, performance, and transparency of our framework under realistic conditions. As mentioned it's keylogging relies on a stealthy polling approach: it repeatedly calls the Windows API GetAsyncKeyState to check the state of each key, rather than installing a hook via the OS. This design makes it stealthier in some respects and is fairly typical of many simple yet effective keyloggers in the wild.

\textbf{Attacker’s Technique:} WarzoneRAT’s keylogger uses a polling loop that repeatedly calls \texttt{GetAsyncKeyState} across virtual key codes (0x08–0xFE) to detect pressed keys. It supplements this with calls to \texttt{GetForegroundWindow}, \texttt{GetWindowTextW}, and \texttt{GetKeyboardState} to capture the active window title and modifier keys (e.g., Shift, CapsLock), allowing it to reconstruct accurate, timestamped keystroke logs tagged by application context. These logs are periodically written to disk, building a detailed timeline of user activity such as: “[Browser – 10:30AM] username: Alice” followed by “[Browser – 10:30AM] password: secret123”.

\textbf{Deception Deployment:} We evaluated our framework by injecting a hook-enabled DLL into the WarzoneRAT process at runtime. For this experiment, we used the EasyHook-based implementation. Our focus was on subverting WarzoneRAT’s polling-based keylogger, which repeatedly invokes \texttt{GetAsyncKeyState} to collect keystrokes. We installed an inline hook on \texttt{GetAsyncKeyState} and redirected all invocations to a custom detour handler governed by our \textit{Decoy Injection} policy. Rather than attempting full behavioral emulation or adaptive input generation, our approach focuses on direct substitution of keystroke content with carefully crafted, plausibly structured decoy sequences. During evaluation, the detour handler replaced the output of \texttt{GetAsyncKeyState} with a predefined but realistic-looking credential string—such as \texttt{"hrSmith2025!"}—designed to mimic user input patterns commonly observed in authentication workflows. The injection sequence also included synthetic tab and enter keystrokes to simulate form navigation and submission.  This substitution-based deception simplifies runtime complexity while still producing high-fidelity artifacts that closely resemble genuine user behavior from the attacker’s perspective.

\textbf{Results and Impact:}  
During the evaluation, our framework successfully deceived WarzoneRAT without disrupting its execution or revealing the presence of instrumentation. The malware continued to invoke \texttt{GetAsyncKeyState} at its normal polling frequency, unaware that our hook was intercepting the calls. When triggered during a simulated login scenario, the detour returned a predefined but fake decoy sequence (e.g., \texttt{"hrSmith2025!"}), including synthetic \texttt{TAB} and \texttt{ENTER} keystrokes to mimic form submission. The actual user input remained fully hidden. The resulting keylog file (\texttt{keylog.log}) contained only the decoy data, complete with accurate window titles and timestamps, producing output that appeared indistinguishable from genuine user behavior. The malware’s auxiliary functions—such as modifier key checks and file I/O—proceeded normally, and no detection logic was triggered. To measure runtime performance impact, we used the Windows Performance Monitor (\texttt{perfmon}) to track CPU utilization of the instrumented process during high-frequency polling sessions. Simulated user interactions were performed repeatedly over a 60-second interval, with the keylogger running under normal conditions. Across five independent runs, the measured CPU overhead introduced by our hooking and detour logic remained consistently below one percent. The host system remained fully responsive, and no input lag or application slowdown was observed during these tests.

\textbf{Overhead Across Deception Policies:}
We also examined the runtime overhead introduced by our two deception policies: Decoy Injection and Input Perturbation. For static content theft (e.g., clipboard or credential files), Decoy Injection imposes negligible overhead, as it substitutes predefined content without parsing. In contrast, Input Perturbation requires runtime processing to apply masking ratios, increasing computational cost. In polling-based keyloggers like WarzoneRAT, which calls \texttt{GetAsyncKeyState} approximately 131 times per minute, full Decoy Injection requires modifying all 131 calls. However, applying a 20 percent perturbation ratio reduces this to only 26 modified responses, lowering CPU load while still corrupting the keylog. This tradeoff allows our framework to balance deception fidelity with performance efficiency based on the threat model.


\subsection{Broader Evaluation Against Real-World Malware}\label{evalA}

To generalize beyond a single malware instance, we tested our framework against fifty keylogger samples as mentioned in Section \ref{api_coverage}. During our analysis, we observed that several malware families including Agent Tesla, Lokibot, AppleSeed, Azorult, and Astaroth employed basic sandbox and virtualization evasion techniques. These behaviors involved checks for the presence of debuggers, timing anomalies, or signs of an analysis environment using APIs such as \texttt{IsDebuggerPresent}, \texttt{GetTickCount}, and \texttt{NtQueryInformationProcess}. When successful, these checks would cause the malware to suppress its payload execution or terminate early in order to avoid analysis or detection. To prevent these evasive techniques from interfering with full execution, we extended our hooking layer to intercept the relevant evasion-related APIs. Our detour logic returned spoofed values that emulated a normal, non-instrumented environment. For instance, debugger detection APIs were forced to return false, and timing outputs were adjusted to eliminate inconsistencies. As a result, the malware continued its execution path and reached the keylogging stage without detecting the instrumentation.

For each sample, we confirmed the malware attempted to invoke one or more input collection APIs listed in Table~\ref{tab:keyloggers}, and verified our framework intercepted them correctly. In all fifty cases, the Hooking Layer captured the targeted API calls and redirected them to the Deception Engine, which returned either crafted decoy data or perturbed input. The malware executed normally, recorded fake inputs, and never logged real user data. None of the samples triggered crashes or evasion logic in response to our instrumentation. This evaluation confirms that our framework reliably deceives a broad range of keylogger families, even when facing common evasion techniques. We next evaluate its resilience against active tampering and anti-hooking defenses.

\subsection{Case Studies: Against Anti-Hooking Attacks}
To rigorously evaluate our framework’s resilience, we developed a custom “super keylogger” that combines multiple input-logging methods with advanced anti-hooking techniques. This case study simulates a sophisticated attacker using diverse strategies while actively evading defensive hooks. Building it ourselves gave us full control to observe the interplay between attack and defense. The process also drew on real-world malware behaviors and AI-assisted development to ensure it was both comprehensive and realistic.

\textbf{Building a Multifaceted Keylogger:} We used an LLM (GPT-4o) to assist in generating and refining keylogger components, which accelerated the implementation of multiple logging methods and ensured comprehensive coverage. Through iterative prompts (e.g., ``generate code to log keys via GetAsyncKeyState," ``what other Windows APIs capture keystrokes?"), we built a modular keylogger integrating five distinct data-capture techniques: (1) direct polling via GetAsyncKeyState and GetKeyState, (2) event hooking through SetWindowsHookEx, (3) GUI message spying using functions like PeekMessage, (4) clipboard monitoring via OpenClipboard and GetClipboardData, and (5) network interception of HTTP POSTs by hooking HttpSendRequest and WSASend, simulating form grabbing. These methods reflect the breadth of userland keylogging seen in real malware.

We implemented each method in its own thread, allowing the keylogger to simultaneously poll keys, monitor clipboard activity, and intercept network traffic—maximizing redundancy and data capture. In tests on Windows 10 and 11, the keylogger reliably recorded all expected input, including clipboard changes and mock web submissions, even producing overlapping logs across techniques. Functionally, it behaved like a super keyloggers—merging traditional and advanced techniques into a single, potent data-theft tool. This confirmed that our deception targets were up against a realistic and comprehensive threat. All prompts, source code, and modular implementations used to construct this keylogger are publicly available in our supplementary materials~\cite{keylogger_techniques}.

\textbf{Integrating Anti-Hooking Techniques and Evaluating Framework Resilience:}\label{superAndevasion}
Next, we equipped the keylogger with four anti-hooking strategies~\cite{smokeloader2,0x00sec-defeating-hooks, cano_reloadlibrary, apostolopoulos2021resurrecting} to simulate an attacker actively trying to evade hook-based defenses like ours. One technique was inspired by the real-world malware SmokeLoader\cite{smokeloader2}, known for its aggressive anti-hooking behavior. Another was derived from academic literature, reflecting techniques commonly discussed in research\cite{apostolopoulos2021resurrecting}. The remaining two \cite{0x00sec-defeating-hooks, cano_reloadlibrary} were adapted from open-source implementations available on GitHub, mimicking what a skilled attacker could easily integrate. Together, these made the keylogger an actively evasive threat capable of challenging advanced hook defenses.


\subsubsection{Runtime Unhooking via .text Restoration}

This technique involves the keylogger actively removing inline hooks by restoring the original bytes of targeted API functions in memory\cite{smokeloader2}. For example, it marks the memory of \texttt{GetAsyncKeyState} as writable using \texttt{VirtualProtect}, then restores the original bytes—either previously saved before the hook was installed or retrieved from a clean copy of the DLL. If our framework had patched a \texttt{JMP} instruction at the beginning of \texttt{GetAsyncKeyState}, the malware would overwrite it with the original opcodes. This is a direct assault on inline hooks; if successful, control flow no longer jumps to our detour and instead executes the API normally. We have observed real-world malware using this tactic to disable userland security monitors.

\textbf{Outcome:} We anticipated this to be a substantial challenge, as it targets the core of our Hooking Layer. However, the Hook Integrity Manager’s active verification loop responded almost instantly. As soon as the keylogger overwrote the first few bytes, the watchdog thread’s subsequent hash check flagged the discrepancy. Our framework immediately re-patched the hook in the function’s prologue. From the keylogger’s perspective, it had successfully unhooked the API and proceeded without further checks, unaware that within milliseconds, the hook was reinstated. Since the keylogger didn’t continuously verify the hook status, it assumed success and continued calling \texttt{GetAsyncKeyState}, all while our detour fed it synthetic keystrokes. We confirmed this by inspecting its logs—only decoy inputs such as \texttt{"Admin123!"} were recorded post-unhooking. The keylogger remained unaware of the deception, validating our rapid detection and self-healing hook strategy as a defense against memory patching attacks.

\subsubsection{Import Address Table (IAT) Rebinding to Clean DLLs}
In this technique, the keylogger circumvents hooked APIs by loading a clean copy of a system DLL\cite{cano_reloadlibrary}. Using \texttt{LoadLibraryEx} with specific flags, it loads a fresh instance of \texttt{user32.dll} into a different memory location. This new instance, absent during our original hook injection, contains unpatched functions. The malware then uses \texttt{GetProcAddress} to resolve function addresses (e.g., \texttt{GetAsyncKeyState}) and updates its Import Address Table or local function pointers to reference these clean copies. This enables the malware to call APIs from the unmonitored DLL, bypassing our instrumentation.


\textbf{Outcome:} Our framework countered this using the Hook Integrity Manager’s clone-detection. Since we had hooks placed on Windows loader routines, any new DLL load event triggered inspection. Upon identifying the newly loaded module as a clone of \texttt{user32.dll}, the framework scanned its export table, injected the same hooks, and rewired malware’s function pointers to route through our detours. Consequently, while the malware believed it had sidestepped our monitoring, we had already hooked the new entry points. The keylogger resolved and invoked what it assumed was an unhooked function, but execution flowed straight into our deception logic. Again, the logged keystrokes were decoys. This emphasized the importance of real-time module load monitoring. While it introduces some complexity, our experiments showed the added hooks incurred minimal overhead and ensured effective coverage.

\subsubsection{Manual API Resolution (Alternate DLL Loading without IAT)}
An evolution of IAT rebinding, this method avoids import tables altogether. The malware manually resolves addresses of critical APIs at runtime and invokes them directly through function pointers\cite{apostolopoulos2021resurrecting}. For instance, it loads a clean copy of \texttt{ntdll.dll} or \texttt{user32.dll}, uses \texttt{GetProcAddress} to resolve \texttt{GetAsyncKeyState}, and stores the resulting address. It then calls the function via \texttt{(*funcPtr)()}, bypassing static imports and any hooks placed there.


\textbf{Outcome:} We generalized our clone-module defense by maintaining a registry of detoured functions by name and module base address. When a new DLL is loaded, we hook all matching exports. As a result, the function pointers resolved by the malware point to already-detoured functions. Since manual resolution eliminates future interception opportunities, our response must occur during module load. We optimized our loader hook to apply detours preemptively. We also ensured our trampolines were either position-independent or appropriately adjusted for each base address. During tests, the malware resolved \texttt{GetAsyncKeyState}, stored the address, and called it—unaware that the address already led to our handler. No anomalies were detected by the keylogger, confirming the efficacy of our proactive strategy. While this approach requires hooking multiple module instances, our modular architecture reused detour logic efficiently, maintaining low overhead.

\subsubsection{In-Memory Signature Scanning for Hooks}

In this defensive maneuver, the keylogger scans memory to detect the presence of inline hooks\cite{0x00sec-defeating-hooks}. Many malware strains implement such scanning routines, checking whether the first few bytes of sensitive APIs match known patterns or appear tampered. Our keylogger version scanned APIs like \texttt{GetAsyncKeyState} and \texttt{SetWindowsHookEx} for recognizable signatures, such as near jumps \texttt{(E9)} or indirect jumps. Upon detecting a potential hook, it either avoided calling the function or logged a warning, thereby reducing the chance of interacting with a monitored or manipulated function.

\textbf{Outcome:}
Our defense relied on stealth-by-diversity, implemented through trampoline obfuscation. Our detours avoided conventional byte patterns. For example, a hook stub might begin with a \texttt{PUSH} followed by a \texttt{RET}, deviating from typical \texttt{JMP} signatures. We also introduced random padding and varied hook placement through address layout randomization. During experiments, the keylogger ran its scan before logging began. It failed to detect our hooks and proceeded to log activity—entirely unaware it was capturing only fake keystrokes. This reinforced the importance of remaining covert in deception-based defenses. Once the adversary suspects tampering, they may disengage or adapt. Our obfuscated hooks maintained the attacker’s trust in the environment, ensuring continued operation under false pretenses.

We evaluated our framework against a custom “Super” keylogger implementing .text restoration, IAT rebinding, manual API resolution, and in memory hook detection. Despite efforts to remove hooks, load clean DLLs, resolve APIs manually, and detect in-memory modifications, our framework neutralized each evasion method in real time. The keylogger was limited to logging only decoy inputs, with real user activity unaffected. This test confirmed the resilience of our combined use of inline hooking, integrity validation, and deception.

\section{Related Work}\label{Related}

To the best of our knowledge, this is the first system to combine API hooking for deception with integrated defenses against anti-hooking techniques. Existing deception frameworks often fail under tampering, while anti-hooking solutions generally lack deceptive capabilities. As these approaches address distinct aspects of the problem space, direct empirical comparison is infeasible. Our work bridges this gap through a resilient design.

Deception has long been a strategic defense in cybersecurity, with early methods relying on honeypots, honeytokens, and honeywords to mislead adversaries and collect telemetry~\cite{kyung2017honeyproxy, godakanda2024honey, petrunic2015honeytokens}. These mechanisms have since evolved into dynamic frameworks that leverage moving target defense~\cite{pagnotta2023dolos}, Markov decision processes~\cite{hayatle2013markov}, and game-theoretic planning~\cite{anwar2022honeypot} to adapt deception strategies in real time. Systems such as PhantomFS~\cite{lee2020phantomfs} and others~\cite{jafarian2020deception, duan2018conceal, islam2020email} offer scalable platforms for orchestrating and evaluating deception assets.

In the context of keyloggers, deception has been explored through the injection of false input to mislead or expose surveillance software. Simms et al.\cite{simms2017keylogger} proposed a decoy keyboard mechanism to confuse keyloggers, while Wazid et al.\cite{wazid2013framework} developed a honeypot-based framework that simulates user interaction to trigger malicious behavior. These works focus on misdirection without addressing the advanced evasion capabilities. Complementing deception, a substantial body of work focuses on detection. Nasaka et al.\cite{nasaka2011keystroke} and Al-Husainy\cite{al2008detecting} monitored API usage to flag suspicious behavior. HookTracer~\cite{case2020hooktracer} performed forensic memory analysis to detect hook artifacts, while Muthumanickam et al.~\cite{muthumanickam2015effective} targeted inline and IAT hook tampering in userland processes. However, these approaches also lack defenses against anti-hooking techniques.

More sophisticated adversaries, such as those behind SmokeLoader~\cite{smokeloader2}, actively dismantle or bypass defensive instrumentation. SmokeLoader, for instance, restores original function prologues to remove inline hooks. Other public implementations demonstrate evasive tactics including Import Address Table rebinding and alternate DLL loading~\cite{cano_reloadlibrary}. In-memory hook detection using signature scans, as seen in various malware samples~\cite{0x00sec-defeating-hooks}, further complicates defensive visibility. These strategies highlight a growing asymmetry: while detection improves, malware continues to evolve to evade naive userland monitoring hookings entirely.

\section{Discussion and Conclusion}\label{Conclusion}


\textbf{Effectiveness and Resilience:} Our experience shows that hooking-based deception is both feasible and effective for countering userland keyloggers. In all test scenarios, keyloggers operated under the illusion of capturing real input, while our framework injected fabricated data. For example, when deployed against WarzoneRAT, a real-world RAT with polling-based keylogging, our hook on GetAsyncKeyState redirected keystroke queries to our deception engine, which returned realistic decoys. These were logged by the malware with correct context and timestamps, keeping actual user input secure. User experience was unaffected and system overhead remained negligible as presented in~\ref{evalB}.


To evaluate resilience under a stronger adversarial model, we developed a custom super keylogger combining behaviors from diverse malware families. Guided by malware reports, academic literature, and LLM-assisted synthesis, it simulates diverse keylogging strategies without using multiple samples. The keylogger integrates five input-capture techniques and incorporates four advanced evasion methods: one drawn from real-world malware, one from an academic publication, and two from widely used open-source tools. These include .text section restoration, IAT rebinding, alternate DLL loading, and signature-based memory scanning, as detailed in Section~\ref{superAndevasion}. Despite full knowledge of our defense mechanisms, the adversary was unable to bypass the system. The Hook Integrity Manager mitigated each evasion attempt through proactive hook verification, clone DLL tracking, and randomized trampoline construction. These results validate the soundness and resilience of hooking our strategy. The combination of runtime input randomization and multi-layer tamper resistance significantly increases the complexity of static and behavioral evasion. Successful circumvention would require memory inspection or runtime probing, which contradicts keylogging stealth and proved ineffective in our evaluation.




\textbf{Limitations and Challenges:} Despite promising results, several limitations emerged. First, our  framework operates solely in user space, excluding kernel-level and hardware keyloggers, which bypass the API layer. Extending deception to lower layers would require OS or driver-level support, which we leave for future work. Second, our framework assumes a prior detection event. It is not a standalone detector; rather, it activates once a process is flagged as suspicious by a security analyst or detection tool. In our evaluations, we manually injected the module, but in real deployments, this would need to be automated and tightly integrated with existing detection pipelines. Finally, hooking in adversarial settings introduces engineering complexity. Though EasyHook and MS Detours ease initial development, they lack tamper resilience. We added trampoline obfuscation, clone-module tracking, and memory protections, requiring deep understanding of the Windows memory model and low-level tuning to ensure stability and performance.


\textbf{Lessons for Practitioners:} From this experience, several key lessons emerge for practitioners building deception systems. First, designing for evasion from the start is essential. Hook integrity must be treated as a first-class concern, as adversaries will attempt to tamper with instrumentation. Our framework’s success depended on early integration of tamper detection and restoration. Second, deception only works if the attacker believes it. Static or naive decoys risk exposure. In our framework, plausible fake inputs (e.g., passwords with tab/enter) increased believability. Adding variability or randomization in future versions may further enhance stealth. Third, anti-hooking itself can serve as a detection signal. Our design demonstrates that tampering attempts offer forensic value, and deception frameworks can double as sensors for identifying sophisticated threats. Finally, the balance between stealth and response must be carefully considered. We prioritized transparency by not killing threads or processes even during tampering. However, a higher-assurance environment may require stronger countermeasures. Policy-driven flexibility allows tailored responses per threat context.


In conclusion, this work demonstrates that deception via API hooking is a practical and resilient defense against userland keyloggers. By intercepting and manipulating attacker-visible API calls, and protecting hook integrity, we misled malware without affecting user experience. Our findings show adversary-aware instrumentation and stealthy deception form a lightweight, effective alternative to traditional detection and blocking. In future work, we plan to integrate the framework with detection pipelines for automatic engagement and extend our techniques to other attack vectors.

\bibliographystyle{IEEEtran} 
\bibliography{bibliography}

\end{document}